\documentclass[aps,prd,preprint,showpacs,groupedaddress]{revtex4-1}
\usepackage{amsmath, amsfonts, amssymb, mathrsfs, color, cases, subeqnarray}
\usepackage{graphicx}% Include figure files
\usepackage{bm}% bold math

\begin{document}
%\preprint{APS/123-QED}

\title{The collapse distance of femtosecond pulses in air}
%\title{Collapse distance of femtosecond laser pulses with different temporal durations in air}% Multipole fields of a rotating and collapsing sphere}
\author{Cunliang Ma}%
\author{Wenbin Lin}
 \email{Corresponding author (email: wl@swjtu.edu.cn).}
% \author{Xiexing Qi}
% \author{Ismail A. M. Harran}
%\author{Chunhua Jiang}%
% \email{jiangchunhua@my.swjtu.edu.cn}
\affiliation{Institute of Electromagnetics, Southwest Jiaotong University, Chengdu 610031, China}
\date{\today}

\begin{abstract}
%{\color{blue}The effects of group-velocity-dispersion on the collapse distance of the femtosecond laser pulses in air is studied in detail.}
The conventional semi-empirical formula for collapse distance [Phys. Rev. 179, 862 (1969); Prog. Quant. Electr. 4, 35 (1975)] has been widely used in many applications. However, it is not applicable when the dispersion length is smaller than or has similar order-of-magnitude as the collapse distance. For the ``enough short'' pulses, there exists a threshold for the initial peak power, with which the collapse distance has a maximum value due to the competition between the Kerr self-focusing and the group velocity dispersion. New semi-empirical formulas are obtained for the collapse distance of the pulse with the initial power being less or larger than the threshold, and they can match the numerical simulations gracefully.
\end{abstract}
\pacs{42.65.Jx, 42.65.Tg}
\maketitle

\section{INTRODUCTION}
The ultra-short high intensity laser pulses have many applications such as laser-guided electric discharge~\cite{Tzortzakis0}, terahertz generation~\cite{Johnson0} and remote sensing~\cite{Luo0}.
Such applications need extremely high light powers which can be obtained via the collapse of pulses~\cite{Silberberg0,Kasparian0,Berge0,Grow0}.
%These applications depend critically on the collapse dynamics~\cite{Silberberg0,Kasparian0,Berge0,Grow0}.

For a Gaussian beam, if the initial peak power exceeds the critical power for self-focusing, the beam will undergo collapse until the higher-order processes such as plasma or high-order Kerr effects halt the collapse~\cite{Marburger0,Fibich0,Couairon0}.
In most cases, the collapse distance $L_{c}$ (the propagation length of the self-focusing beam until collapse) can be well approximated by a semi-empirical formula~\cite{Dawes0,Marburger0, Couairon0,Couairon1,Couairon2,Xi0,Xi1}
\begin{eqnarray}
L_{c}\approx L_{c}^{semi}=\frac{0.367\pi n_0 r_0^2/\lambda_0}{\sqrt{[(P_{in}/P_{cr})^{1/2}-0.852]^2-0.0219}}~,
\label{semi}
\end{eqnarray}
where $n_0$ is the refractive index, $r_0$ is the beam width  which is at $1/e^2$ level of intensity, $\lambda_0$ is the laser wavelength in vacuum, $P_{in}$ is the initial pulse's peak power, and $P_{cr}$ is the critical power for self-focusing which can be written as $P_{cr}=3.77\lambda_0^2/8\pi n_0n_2$ with $n_2$ being the Kerr index. However, for the very high powers such as $P_{in}=100P_{cr}$,
experiments and numerical simulations show that the collapse distance can not be described by $L_{c}^{semi}$ and a transition from a $1/\sqrt{P_{in}}$ to a $1/P_{in}$ scaling was observed~\cite{Fibich1}. Recently, simulations show that the group-velocity-dispersion (GVD) has a great influence on $L_{c}$ when the pressure is relatively high, e.g., $10~\textrm{atm}$~\cite{Li0}.

In this paper, we investigate the collapse distance of femtosecond laser pulses in air for different temporal durations.
We find that for the ``very short'' pulses, there exists a threshold for the initial power, with which the collapse distance has a maximum value. More importantly, new semi-empirical formulas are obtained for the collapse distance of the pulse with the initial power being less than or larger than the threshold. The new formulas take into account of the influence of GVD and can match the numerical simulations perfectly, including the cases of high pressures at which the GVD becomes more important. %The paper is organized as follows. Sec. 2 gives a brief introduction for the model of the femtosecond pulse propagation in air. Numerical simulation results are presented in Sec. 3. Summary is given in Sec. 4.

\section{THEORETICAL MODEL}
The physical processes which halt beam collapse is still controversial. Some groups think the plasma prevents collapse (Kerr-plasma model)~\cite{Kolesik0,Kolesik1,Polynkin0,Warand0,Kosareva0,Kohler0,Spott0}, while other groups believe the collapse is stopped by the high-order Kerr effects (HOKE model)~\cite{Bejot0,Ettoumi0,Bejot1,Bree0,Loriot2,Petrarca0,Bejot2}. Fortunately, the collapse distances are almost the same for Kerr-plasma model and HOKE model~\cite{Bejot0,Ma0}, since the HOKE and the plasma effect becomes important only after the pulse has collapsed~\cite{Fibich2}, and this phenomena has also been confirmed in our numerical simulations (not shown here).
In this work we study the collapse distance of the pulse for different temporal durations via HOKE model, which %the propagation of femtosecond laser pulses in air
can be described by an extended non-linear Schr\"odinger equation (NLSE)~\cite{Bejot0,Bejot1}
\begin{eqnarray}
\nonumber\frac{\partial{\sl{A}}}{\partial{\sl{z}}} &=&\frac{\mathrm{i}}{2k_0}\Delta_{\perp}\sl{A}-
\frac{\mathrm{i}k''}{2}\frac{\partial^2{\sl{A}}}{\partial{\sl{\tau}^2}}+\frac{\mathrm{i}k_0}{n_0}
  \left(\sum_{j=1}^{4}n_{2*j}\left|\sl{A}\right|^{2*j}\right)\sl{A}-\\
  &&\frac{\mathrm{i}k_0}{2}\frac{\omega_{pe}^2}{\omega_0^2}\sl{A}
-\frac{\sl{A}}{2}\sum_{l=\mathrm{O_2},\mathrm{N_2}}\left(\frac{W_l(I)U_l}{|A|^2}(\rho_{at,l}-\rho_{e,l})\right),
%\frac{\beta^{(K)}}{2}\left|\sl{E}\right|^{2*K-2}\sl{E}, \label{NLS}\\
%\frac{\partial\rho}{\partial t}&=&\frac{\beta^{(K)}}{K\hbar\omega_0}\left|\sl{E}\right|^{2K}\left(1-\frac{\rho}{\rho_{at}}\right),\label{plasma}
\label{NLSE}
\end{eqnarray}
where $A$ represents the envelope of the electric field, and $z$ denotes the propagation distance. $k_0=2\pi/\lambda_0$ ($\lambda_0=$800 nm) is the central wave number.
 %is the wavelength of the laser beam.
The Laplacian operater  $\Delta_{\perp}=\partial^2_r+1/r\partial_r$ denotes the beam transverse diffraction.
The remaining terms in the right-hand-side of Eq.(1) account for the group velocity dispersion with the second order dispersion coefficient $k''=0.2~\textrm{fs}^2/\textrm{cm}$,
Kerr and high-order Kerr effect with nonlinear refractive index $n_2=1.2\times 10^{-19}~\textrm{cm}^2/\textrm{W}$, $n_4=-1.5\times 10^{-33}~\textrm{cm}^4/\textrm{W}^2$,
$n_6=2.1 \times 10^{-46}~\textrm{cm}^6/\textrm{W}^4$, and $n_8=-0.8\times 10^{-59} ~\textrm{cm}^8/\textrm{W}^4$~\cite{Loriot0,Loriot1}, plasma defocusing with the plasma oscillation frequency $\omega_{pe}=\sqrt{q_e^2\rho/m_e\epsilon_0}$ ($q_e$ is the electron charge, $m_e$ is the electron mass and $\rho$ is the free electron density.),
the energy loss caused by MPA.
$W_{N_2}(I)$ and $W_{O_2}(I)$ are the photoionization rate of N$_{2}$ and O$_2$.
$\rho_{at,N_2}=2.1\times 10^{25}$ m$^{-3}$ and $\rho_{at_,O_2}=5.7\times 10^{24}$ m$^{-3}$ are the density of N$_{2}$ and O$_{2}$ molecular at 1 atm. $\rho_{e,N_2}$ and $\rho_{e,O_2}$ are free electron density ionized by N$_{2}$ and O$_{2}$ ($\rho$=$\rho_{e,N_2}+\rho_{e,O_2}$)
which can be calculated by the following equations
%{\color{red}$\rho$=$\rho_{e,N_2}+\rho_{e,O_2}$ is the total free electron density which satisfy the charge conservation equation that can be written as}
\begin{eqnarray}
\label{roueN}
\frac{\partial\rho_{e,N_2}}{\partial \tau}=W_{N_2}(I)\left(\rho_{at,N_2}-\rho_{e,N_2}\right)~,
\end{eqnarray}
\begin{eqnarray}
\label{roueO}
\frac{\partial\rho_{e,O_2}}{\partial \tau}=W_{O_2}(I)\left(\rho_{at,O_2}-\rho_{e,O_2}\right)~.
\end{eqnarray}
The photoionization rate of N$_2$ and O$_2$ is obtained by the Keldysh-PPT (Perelomov-Popov-Terent'ev) formula~\cite{Kasparian1}.
%We find that for the ``enough short'' pulses, there exists a threshold for the initial power, with which the collapse distance has a maximum value. More importantly, a new semi-empirical formula is obtained for the collapse distance of the pulse with the initial power being larger than the threshold. This new formula takes into account of the influence of group-velocity-dispersion and can match the numerical simulations perfectly, including the cases of high pressures.

The initial pulse investigated in this paper is Gaussian beam which can be written as:
\begin{eqnarray}
\label{gauss}
E(r,\tau,0)=\sqrt{\frac{2P_{in}}{\pi r_0^2}}\exp{\left(-\frac{r^2}{r_0^2}-\frac{\tau^2}{\tau_0^2}\right)}~,
\end{eqnarray}
where $\tau_0$ is the temporal duration.

\section{RESULTS AND DISCUSSIONS}  %This should be cancelled for PRL and APL.
\begin{figure*}[htb]
\includegraphics[width=10.5cm]{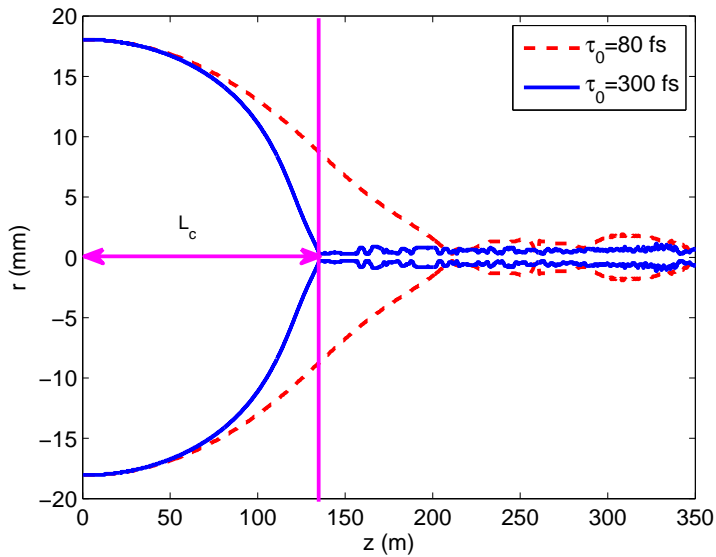}% Here is how to import EPS art
\caption{\label{fig:Figure1}Beam radius of the pulse at different propagation distances.
The input beam are Gaussian pulses with $r_0=18~\textrm{mm}$ and $P_{in}/P_{cr}=20$.}
\end{figure*}

Follow Li et al. ~\cite{Li0}, we define the collapse distance as the distance between the light source and the position where the laser beam has the smallest radius.
Fig. \ref{fig:Figure1} shows the evolution of the beam radius of pulses which have different temporal durations ($80~\textrm{fs}$ and $300~\textrm{fs}$). It can be seen that the collapse distances of the two pulses have big difference ($207~ \textrm{m}$  and $135~\textrm{m}$). In contrast, the semi-empirical formula in Eq.~(\ref{semi}) gives $L_c^{semi}=129~\textrm{m}$ since it does not takes into account of the pulse's duration. Therefore, the temporal duration of pulse may have a large influence on the collapse distance in some circumstances.

%
%\begin{figure}[htb]
%\centerline{\includegraphics[width=10.5cm]{Figure1.eps}}
%\caption{Beam radius of the pulse at different propagation distances.
%The input beam are Gaussian pulses with $r_0$=18 mm, $P_{in}/P_{cr}=20$.\label{fig:Figure1}}
%\end{figure}

From now on, we focus on the influence of the temporal duration to the collapse distance. In the simulations, we consider the Gaussian pulse with two different beam radius ($r_0=2~\textrm{mm}$ and $r_0=9~\textrm{mm}$).

\begin{figure}[htb]
\centerline{\includegraphics[width=10.5cm]{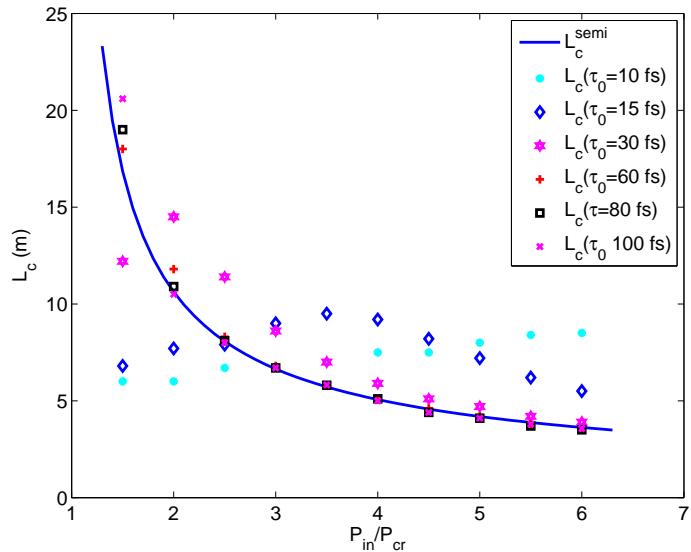}}
\caption{Variation of the collapse distance of pulses with the initial peak power. The input beam are Gaussian pulses with $r_0=2~\textrm{mm}$.
\label{fig:Figure2}}
\end{figure}

\begin{figure}[htb]
\centerline{\includegraphics[width=10.5cm]{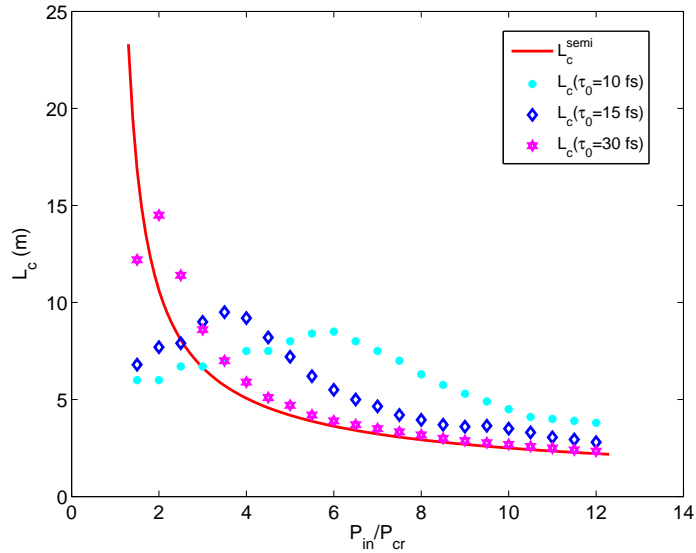}}
\caption{Variation of the collapse distance of the short pulses with the initial peak power. The input beam are Gaussian pulses with $r_0=2~\textrm{mm}$.
\label{fig:Figure3}}
\end{figure}

Fig. \ref{fig:Figure2} shows the variation of the collapse distances with different initial powers and temporal durations.
Here $r_0$=2 mm, $P_{in}/P_{cr}$ varies from 1.5 to 6, and $\tau_0$ varies from 10 to $100~\textrm{fs}$.
For comparisons, the collapse distances calculated by $L_c^{semi}$ is also given and shown in solid line.
We can observe that $L_c^{semi}$ can approximate $L_c$ very well for the long durations such as $\tau_0$=60fs, 80 fs and 100 fs, but it fails for the short ones, e.g., $\tau_0$=$10~\textrm{fs}$, $15~\textrm{fs}$ and $30~\textrm{fs}$. In order to see more clearly how the collapse distances of the short-temporal-duration pulses change with the input power, we show the results in Fig. \ref{fig:Figure3} for $\tau_0$=$10~\textrm{fs}$, $15~\textrm{fs}$, $30~\textrm{fs}$ and $P_{in}/P_{cr}$ being in the range between 1.5 and 12. It can be seen from this figure that the collapse distances of pulses with these short durations have a maxima and tend to $L_c^{semi}$ in the limit of $P_{in}/P_{cr}$ going to infinity. We define the input peak power which leads to the maximum collapse distance as $P_{MC}$. If $P_{in} < P_{MC}$, the collapse distance increases with $P_{in}$, and if $P_{in} > P_{MC}$, the collapse distance decreases with $P_{in}$.


\begin{figure}[htb]
\centerline{\includegraphics[width=10.5cm]{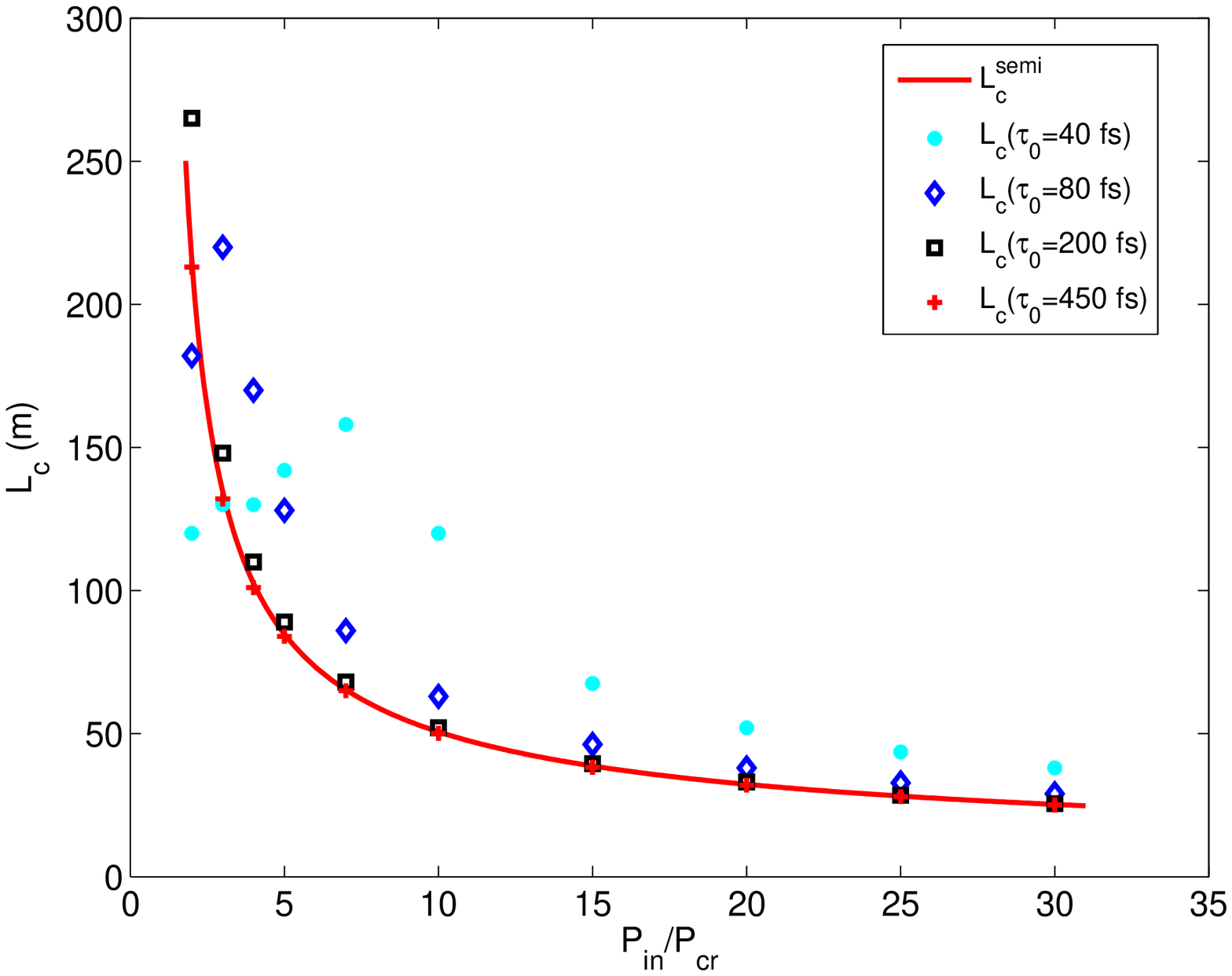}}
\caption{Variation of the collapse distance with the initial peak power ($P_{in}/P_{cr}$ range from 2 to 30). The input beam are Gaussian pulses with $r_0=9~\textrm{mm}$. The temporal durations of the pulses are $40~\textrm{fs}$, $80~\textrm{fs}$, $200~\textrm{fs}$, $450~\textrm{fs}$.
\label{fig:Figure4}}
\end{figure}

Fig. \ref{fig:Figure4} shows the collapse distances for the pulses with $r_0$= 9 mm. $P_{in}/P_{cr}$ varies from 2 to 30, and $\tau_0=$$40~\textrm{fs}$, $80~\textrm{fs}$, $200~\textrm{fs}$, $450~\textrm{fs}$, respectively.
From Fig. \ref{fig:Figure4} we can get the same conclusion as shown in Fig. \ref{fig:Figure2} and \ref{fig:Figure3} that $L_c$ has a maximal value when the temporal duration is very short.
It is worth pointing out that ``very short'' is relative to the radius of pulses. For example, $\tau_0$= 80 fs is ``very short'' when $r_0$=$9~\textrm{mm}$, whereas it is not ``very short'' when $r_0$=$2~\textrm{mm}$.

The phenomena that $L_c^{semi}$ can not fit $L_c$ well for the ``very short'' pulses is due to GVD. The normal GVD enables power exchanges between different time slices of the pulse and disperses the latter in time, and this may contribute to maintaining the pulse self-guiding at smaller intensity levels~\cite{Berge2,Champeaux0,Champeaux1}.
If the pulse's temporal duration is long enough, the effects of GVD can be neglected when the propagation distance is less than the collapse distance, and thus $L_c^{semi}$ can fit the collapse distance very well. However,
if the pulse's temporal duration is very short, the GVD effect will have a strong competition to the Kerr self-focusing effect when the initial power is small. Let's consider an extremely short pulse with small initial peak power.
In this extreme situation, the GVD effect overwhelms the Kerr self-focusing, and the collapse distance equals the propagation distance at which the peak intensity decreases to $P_{cr}$.
At the same time, the temporal duration varies with $z$ as $\tau_1(z)=\tau_0[1+(z/2L_{GVD})^2]^{1/2}$~\cite{Agrawal0}, here $L_{GVD}=\tau_0^2/2k''$ denotes the dispersion length~\cite{Couairon1}, and thus the peak power can be written as $P(z)=P_{in}/\tau_1(z)$. The collapse distance in this case can be approximated by
\begin{equation}
\label{extreme}
L_c^{extreme}=2L_{GVD}\sqrt{(P_{in}/P_{cr})^2-1}~.
\end{equation}
This equation may account for the phenomena that the collapse distances of short pulses increase with $P_{in}/P_{cr}$ when $P_{in}<P_{MC}$ (Figs. \ref{fig:Figure2}-\ref{fig:Figure4}), though Eq. \ref{extreme} can not be used to estimate $L_c$ accurately in general cases.

With closer inspections for Figs. \ref{fig:Figure2}-\ref{fig:Figure4}, we find that $L_{c}$ is proportional to $P_{in}/P_{cr}$ when $P_{in} < P_{MC}$, and the slope is dependent on the temporal duration and the radius of pulse. Taking into account the GVD effects, we finally obtain the new semi-empirical formulas as follow %for $P_{in}<P_{MC}$ as follow
\begin{subnumcases}
{L_{c}^{semi,new}=}
%(0.22\pm 0.03) L_{GVD}\frac{P_{in}}{P_{cr}}+(0.32 \pm 0.02) L_{DF}~,~~~~ & \textrm{if~~$P_{in}<P_{MC}$}~,\label{NEW1}\\
0.225 L_{GVD}\frac{P_{in}}{P_{cr}}+0.320 L_{DF}~,~~~~ & \textrm{if~~$P_{in}<P_{MC}$}~,\label{NEW1}\\
L_{c}^{semi}\times 1.878^{N}~, ~~~~& \textrm{otherwise}~,\label{NEW2}
\end{subnumcases}
where $L_{DF}=\pi n_0r_0^2/\lambda_0$ is the Rayleigh length, which account for the contribution from the diffraction. $N$ is a dimensionless parameter and
\begin{eqnarray}
N=\frac{L_c^{semi}}{L_{GVD}}~,
\end{eqnarray}
which characterizes the relative importance of the Kerr self-focusing and the GVD. When $N\gg 1$ GVD dominates, while for $N\ll 1$ Kerr self-focusing dominates.

\begin{figure}[htb]
\centerline{\includegraphics[width=10.5cm]{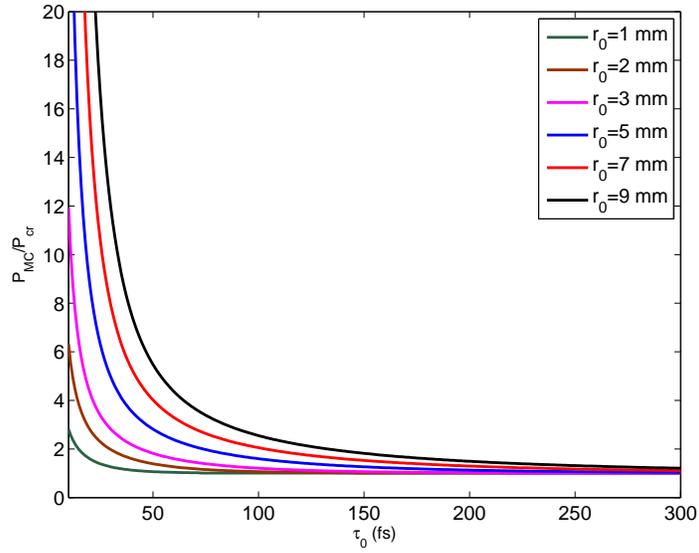}}
%\centerline{\includegraphics[width=10.5cm]{Figure62.eps}}
\caption{
Dependence of $P_{MC}$ on the pulse's duration and pulse's radius.
\label{fig:Figure5}}
\end{figure}

From Eq. (7), we can see that when $P_{in}<P_{MC}$ the collapse distance is mainly determined by the GVD and the diffraction, otherwise, all the GVD, the diffraction and the Kerr self-focusing plays important roles. The input power threshold $P_{MC}$ can be obtained via equaling Eq. (\ref{NEW1}) and Eq. (\ref{NEW2}). The relations between $P_{MC}$, the duration $\tau_0$ and the radius $r_0$ of the laser pulse are shown in Fig. \ref{fig:Figure5}. It can be seen that $P_{MC}$ increases with $r_0$. At the same time, $P_{MC}$ decreases with $\tau_0$, and it reduces to $P_{cr}$ in the the limit of $\tau_0 \to \infty$.
%Therefore, $P_{MC}$ may be regarded as the new $P_{cr}$ when the GVD is taken into consideration. For the variation of the initial peak power there are three regime that control the propagation of the laser beam. {\color{red}When $P_{in}<P_{cr}$ the beam radius increase with the propagation distance and the intensity decrease with the propagation distance. When $P_{cr}<P_{in}<P_{MC}$ Kerr self-focusing overcomes diffraction but the intensity increased by self-focusing may can not cancel out it decreased by the GVD. When $P_{in}>P_{MC}$ the intensity increased by self-focusing can overcome it decreased by GVD.}

In real applications of the new semi-empirical formula, we do not need the value of $P_{MC}$. Notice that Eq. (\ref{NEW1}) is a monotonically increasing function of $P_{in}$, while Eq. (\ref{NEW2}) is a monotonically decreasing function of $P_{in}$, and the value of input power corresponding to the intersection of these two functions is $P_{MC}$. Therefore, we can directly use $\textrm{min}\{\textrm{Eq.}~(\ref{NEW1}),~\textrm{Eq.}~(\ref{NEW2})\}$ to obtain the collapse distance $L_c$ for a given $P_{in}$, without knowing the value of $P_{MC}$.

\begin{figure}[htb]
\centerline{\includegraphics[width=10.5cm]{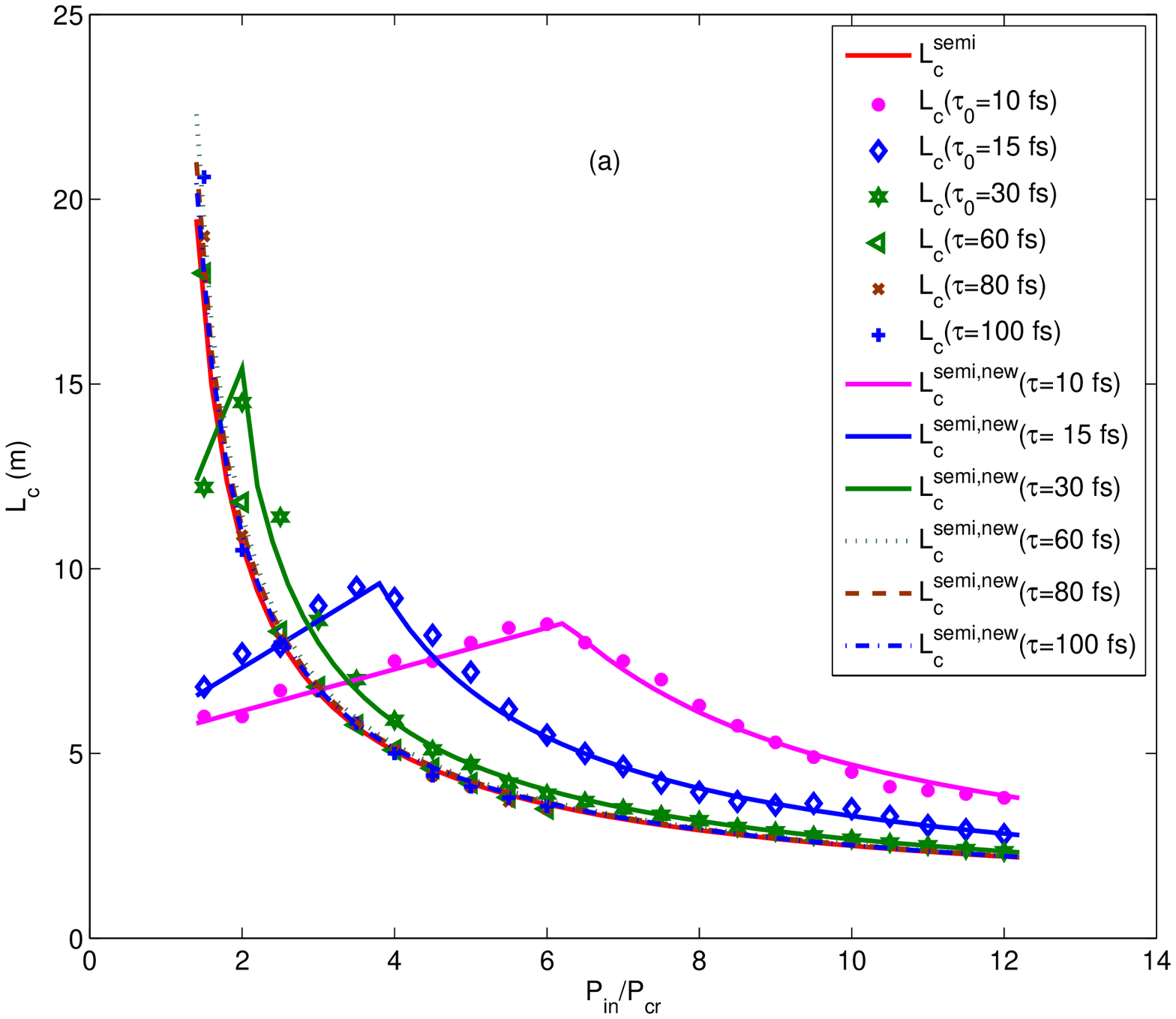}}
\centerline{\includegraphics[width=10.5cm]{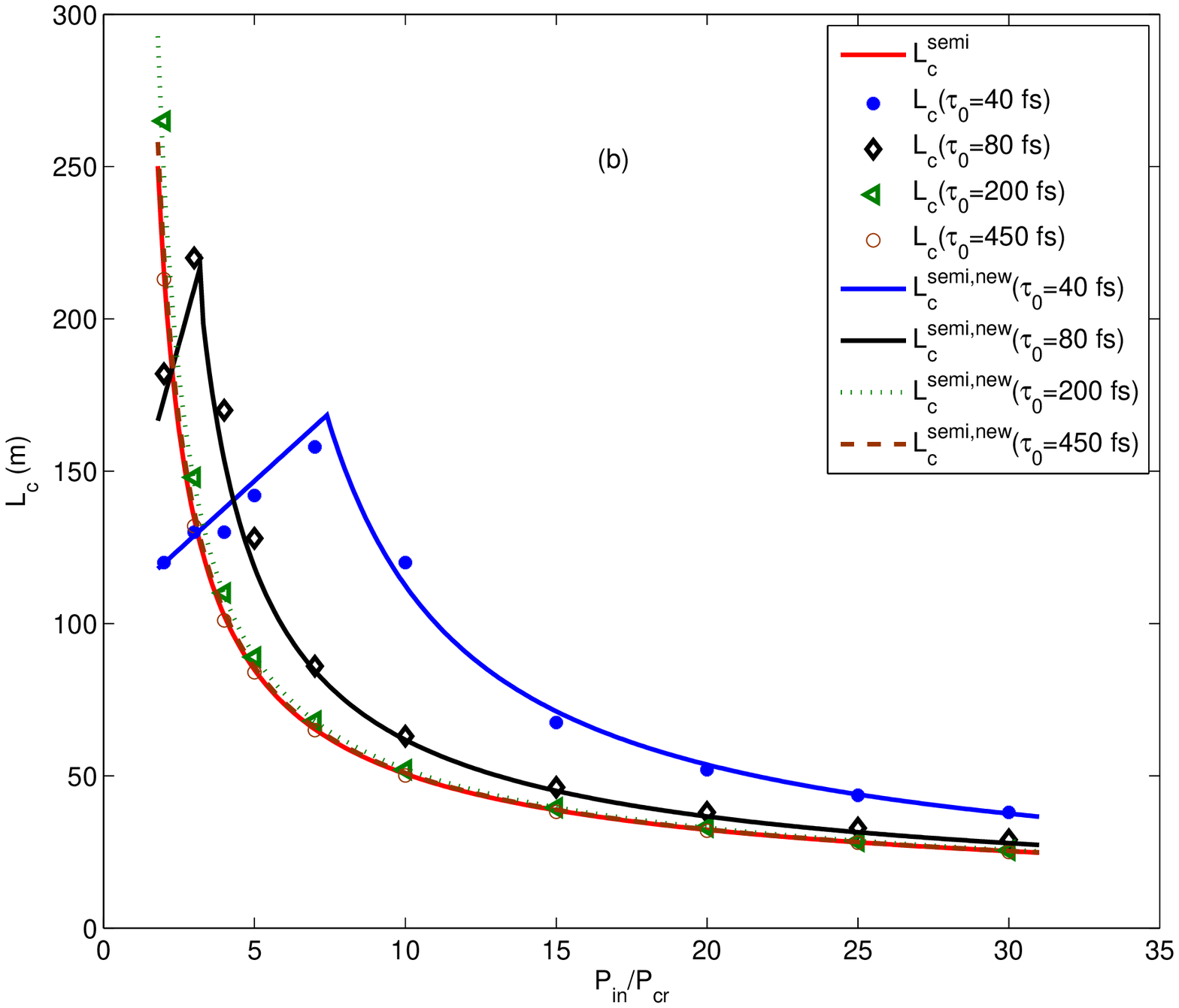}}
\caption{
The collapse distance predicted by the new semi-empirical formula. The input beam are Gaussian pulses with (a) $r_0$=2 mm, (b) $r_0$=9 mm.
%Variation of $P_{MT}$ with the input tempora. The input beam is a Gaussian with $r_0$=2 mm.
\label{fig:Figure6}}
\end{figure}

Fig. \ref{fig:Figure6} shows the comparisons of the collapse distances between the new semi-empirical formula, the direct numerical simulations, and the conventional semi-empirical formula. It can be seen that the new semi-empirical formula agree with the simulation results very well, for both the long and short pulses.
%From Fig. \ref{fig:Figure6} we can see that when the pulse's temporal length is long enough $L_c^{semi,new}$ approximate the $L_c^{semi}$. The new semi-empirical formula is applicable to describe $L_c$ for both the ``very short'' and the long pulses.

%\begin{figure}[htb]
%\centerline{\includegraphics[width=10.5cm]{Figure5.eps}}
%\caption{Variation of $P_{MT}$ with the initial temporal duration. The initial beam are Gaussian pulses with $r_0$=2 mm.
%\label{fig:Figure5}}
%\end{figure}

Now we turn to the case shown in Fig.~1, in which the collapse distances are obtained by numerical simulations. Basing on the new semi-empirical formula given in Eq. (7), we obtain the collapse distance of $214~\textrm{m}$ (207 m in simulation) for $80~\textrm{fs}$ and 134 m (135 m in simulation) for $300~\textrm{fs}$.

Eq. (7) can also be used to calculate the collapse distance at high pressures. %, which Eq. (\ref{semi}) will not be applicable~\cite{Li0}.
%The phenomenon observed by Li et. al.~\cite{Li0} can be explained by Eq. (\ref{NEW}).
$L_{GVD}$ is inversely proportional to the pressure, and play more important roles at high pressure. For example, in the work of ~\cite{Li0}, the authors obtained the collapse distance of the pulse with $\tau_0=50$ fs and $r_0=1.2$~mm, $P_{in}=4P_{cr}$ is 2.3 m at 10 atm, while the new semi-empirical formula gives $L_c=2.2$ m, perfectly matched.

\section{CONCLUSION}
GVD plays important roles in the evolution of the laser pulses in air. For the short pulses, due to the competition between GVD and Kerr self-focusing, there exists a threshold $P_{MC}$ for the initial power, with which the collapse distance has a maximum value. If the initial power is less than the threshold, the collapse distance is proportional to the input power. If the initial power is larger than the threshold, the collapse distance decreases with the the input power, and is larger than that obtained by the conventional semi-empirical formula. Taking into account the effects of GVD, we present new semi-empirical formulas, which match the numerical simulations very well. The formulas can also be applicable to the cases for high pressures in which the GVD effects become more important.
\section{ACKNOWLEDGMENTS}
This work was supported in part by the Ph.D. Programs Foundation of Ministry of Education of China Grant No. 20110184110016, the National Basic Research Program of China (973 Program) Grant No. 2013CB328904, and the Fundamental Research Funds for the Central Universities.

\end{document}